\documentclass[conference]{IEEEtran}
\IEEEoverridecommandlockouts                              

\usepackage{graphics} 
\usepackage{graphicx}
\usepackage{array}
\usepackage{caption}
\usepackage{algorithm}
\usepackage{algorithmicx} 
\usepackage{algpseudocode}
\usepackage{amsmath}  
\usepackage{multirow}
\usepackage{cite}
\title{
\vspace{-0pt} Design Flow of Accelerating Hybrid Extremely Low Bit-width Neural Network in Embedded FPGA \vspace{-4pt}}

\author{Junsong Wang$^{1}$, Qiuwen Lou$^{2}$, Xiaofan Zhang$^{3}$, Chao Zhu$^{1}$, Yonghua Lin$^{1}$, Deming Chen$^{3}$\\ 
$^{1}$IBM Research-China, Beijing, $^{2}$University of Notre Dame\\
$^{3}$University of Illinois at Urbana-Champaign\\
\textit{\{junsongw, bjzhuc, linyh\}@cn.ibm.com, qlou@nd.edu, \{xiaofan3, dchen\}@illinois.edu}
\vspace{-16pt}
}

\begin{document}

\maketitle
\thispagestyle{empty}
\pagestyle{empty}

\begin{abstract}

Neural network accelerators with low latency and low energy consumption are desirable for edge computing. To create such accelerators, we propose a design flow for accelerating the extremely low bit-width neural network (ELB-NN) in embedded FPGAs with hybrid quantization schemes. This flow covers both network training and FPGA-based network deployment, which facilitates the design space exploration and simplifies the tradeoff between network accuracy and computation efficiency. Using this flow helps hardware designers to deliver a network accelerator in edge devices under strict resource and power constraints. We present the proposed flow by supporting hybrid ELB settings within a neural network. Results show that our design can deliver very high performance peaking at 10.3 TOPS and classify up to 325.3 image/s/watt while running large-scale neural networks for less than 5W using embedded FPGA. To the best of our knowledge, it is the most energy efficient solution in comparison to GPU or other FPGA implementations reported so far in the literature.

\end{abstract}

\vspace{+0pt}
\section{Introduction}
With more advanced Deep Neural Networks (DNNs) developed every day, there is a big challenge to deploy such complicated models in edge devices under strict constraints on power budget, delay, form factor, and cost. Researchers are actively investigating approaches to map the DNN workloads onto FPGAs by taking advantage of their improved latency and energy efficiency in order to deliver intelligence from edge devices \cite{zhang2017machine,zhao2017accelerating,zhuge2018face,guan2017fp,umuroglu2017finn}.
One of the most arresting approaches is to reduce the complexities and model sizes of DNNs. Existing works on neural network model compression can be divided into two main approaches: (1) compress the original large and complicated network into a smaller one, or (2) 
directly generate a network with extremely low bit-width (ELB) data. 
Deep compression \cite{han2015deep_compression} is a representative work of the first approach, which uses weak connection pruning, weight quantization, and Huffman coding to compress the network model and remove the redundant network connections. In the second approach, Binarized Neural 
Networks (BNN) \cite{Courbariaux-binary} and Ternary Neural Network (TNN) \cite{Ternarynet} are proposed to constrain the weights to binary and ternary bit-width respectively. 
These two approaches can both significantly reduce the computation complexity and network model size, while their corresponding acceleration designs are discussed in \cite{han2016eie} and \cite{Nurvitadhi2016AcceleratingBN}, respectively.

In this work, we introduce more flexibility on bitwidth optimization compared to previous works, e.g., BNN, TNN and the DoReFa-Net \cite{dorefa}, and adopt a hybrid quantization scheme for both weights and activations so that different bit-widths can be applied for both inter- and intra-layers. 
For instance, our proposed design method allows us to use binary weights in fully-connected (FC) layers for significantly cutting down their demands of memory access bandwidth while use ternary weights for the convolutional (CONV) layers to balance the network accuracy and computational overhead. In terms of the first and the last layer, where usually higher data precision is required, we would use more bit-widths instead of sticking with the low precision options. Such flexible quantization schemes are helpful for building highly customized DNN accelerators in hardware with high capacity to trade off the network accuracy and resource utilization.

Although GPUs are widely deployed for accelerating the DNN training and inference, they are not very efficient for handling the hybrid ELB-NNs with non-uniform quantization schemes. On the contrary, FPGA based solution is more suitable to address the arbitrary numerical precisions by taking advantage of its fine-grained customization capability, and the abundant LUT resource can also be effectively utilized for computation.


We will present a design flow of accelerating large scale hybrid ELB-NN deployment in FPGA for edge devices. Using flexible quantization schemes offers better balance among accuracy, latency, and throughput when we target embedded FPGAs in edge devices. 
We develop a Caffe based framework for efficient training and accuracy evaluation of ELB-NN. To the best of our knowledge, it is the {\bf first} ELB network design flow that enables different bit-width quantizations for inter- and intra-layers in a comprehensive manner. In order to accommodate the diverse quantization schemes in the proposed ELB-NN design method and seek for higher throughput, we construct a pipeline architecture for boosting the inference of DNNs. To make ELB-NN on FPGA implementation much easier, we build an end-to-end automation tool, called AccELB. Given a hybrid ELB-NN model, AccELB can automatically generate a high performance FPGA implementation with pre-built optimized RTL libraries.
To summarize, the detailed contributions of this work are:
\vspace{-0pt}
\leftmargini=0mm
\leftmarginii=0mm
\begin{itemize}
\item \textbf{Hybrid ELB-NN for accuracy and computational complexity tradeoffs}. 
Our experimental results indicate that the accuracy varies with the precisions of weights and activations with regard to the layer types (e.g. CONV, FC) and the layer locations. We observe that the precision of the activation function is in general more sensitive than that of the weights. Also, the first and the last layers require higher data precision than other layers. Our proposed flow allows network designers to adjust the precisions of weights and activations within or across the layers, which provides more flexibility of effective accuracy and throughput/energy trade-offs. 


\item \textbf{An end-to-end automation tool}. To easily deploy the hybrid ELB-NN in embedded FPGA, we develop an integrated automation tool -- AccELB, to map hybrid ELB-NN definitions (e.g. the Caffe prototxts) to FPGA board-level implementations. With such an integrated automation tool, users are no longer required to program at RTL level nor to perform manual design space exploration. 

\end{itemize}

The rest of this paper is organized as follows. In Section 2, the background and related works are introduced. In Section 3, the whole design flow from training to deployment is described. The hybrid ELB-NN and the automation tool AccELB are described in Section 4 and 5 respectively. Experimental results are provided in Section 6. Section 7 concludes this paper.

\section{Background and related work}
\subsection{Low Precision Neural Network}
\vspace{-0pt}
Recent research works have considerably reduced the DNN model size and computation complexity by quantizing the weights and activations to low bit-widths while retaining similar accuracy \cite{Courbariaux-low-precison}\cite{gysel2016hardware}. A much more aggressive quantization approach was proposed in \cite{Courbariaux-binary} that binarizes both the weights and activation to +1/-1, but it only achieved good performance for some small datasets like MNIST, CIFAR10 and SVHN. XNOR-Net by Rastegari \cite{xnor-net} extended the binary nerual network (BNN) to large scale ImageNet dataset and achieved competitive top-1 accuracy for both fully or partially binarized network. DoReFa-Net \cite{dorefa} investigated the effects of the low precision not only in inference but also in training stage by quantizing the gradients in back propagation. Ternary weight networks \cite{Ternarynet}\cite{ternary}, with weights constrained to +1, 0 and -1 along with a scaling factor, were proposed to increase the capacity of model expression, and achieved approximated accuracy of full precision weight networks. In \cite{ADMM}, the authors adopted the Alternating Direction Method of Multipliers (ADMM) to train the network with extremely low bit weights, and achieved state-of-the-art performance.   

\vspace{-0pt}
\subsection{Deep Neural Network Acceleration on FPGA}
\vspace{-0pt}
There are many studies on accelerating the neural networks in FPGA. The authors in \cite{jason15fpga} proposed an analytical design scheme for FPGA with a roofline model and used optimization techniques such as loop tiling and transformation to deal with the memory bandwidth issue. In \cite{wangyu16fpga}, the authors further studied larger networks and proposed dynamic-precision data quantization method and a convolver design to improve the bandwidth and resource utilization. The authors in \cite{2017High} proposed a resource allocation scheme for minimizing the network inference latency and deployed different quantization schemes for the weights and feature maps. Previous literature also focused on building framework for auto-generating FPGA-based DNN accelerator \cite{guan2017fp}.

There are also studies of mapping the low bit-width networks onto FPGA targeting embedded systems. In \cite{zhao2017accelerating}, a neural network with binarized weights was implemented on a tiny FPGA using HLS by caching all the feature map data in the on-chip memory, and demonstrated its efficiency for networks with small model size. FINN in \cite{umuroglu2017finn} introduced methods to optimize the layers in a binary network to meet throughput targets. A detailed comparison of implementing BNN in CPU, GPU, FPGA and ASIC was conducted in \cite{Nurvitadhi2016AcceleratingBN}. Overall, these previous works only target small or full binary neural networks. 
The systematic study for hybrid low bitwidth versus accuracy tradeoffs across different intra- and inter-layer scenarios is also not provided. 

\section{Design Flow of Hybrid ELB-NN}
\label{deign_overview}
In this section, we investigate the hybrid quantization scheme, since different precisions of the weights and activations within or across the layers may have significantly different impacts on the accuracy and throughput. The relationship is non-linear and subtle (i.e., reduction of bitwidths might not necessarily reduce the accuracy in a linear fashion), and our work attempts to provide insights and principles studying the relationship across these different design dimensions. 

\begin{figure}[t]
  \centering
  \setlength{\belowcaptionskip}{-8pt}
  \includegraphics[width=0.98\columnwidth]{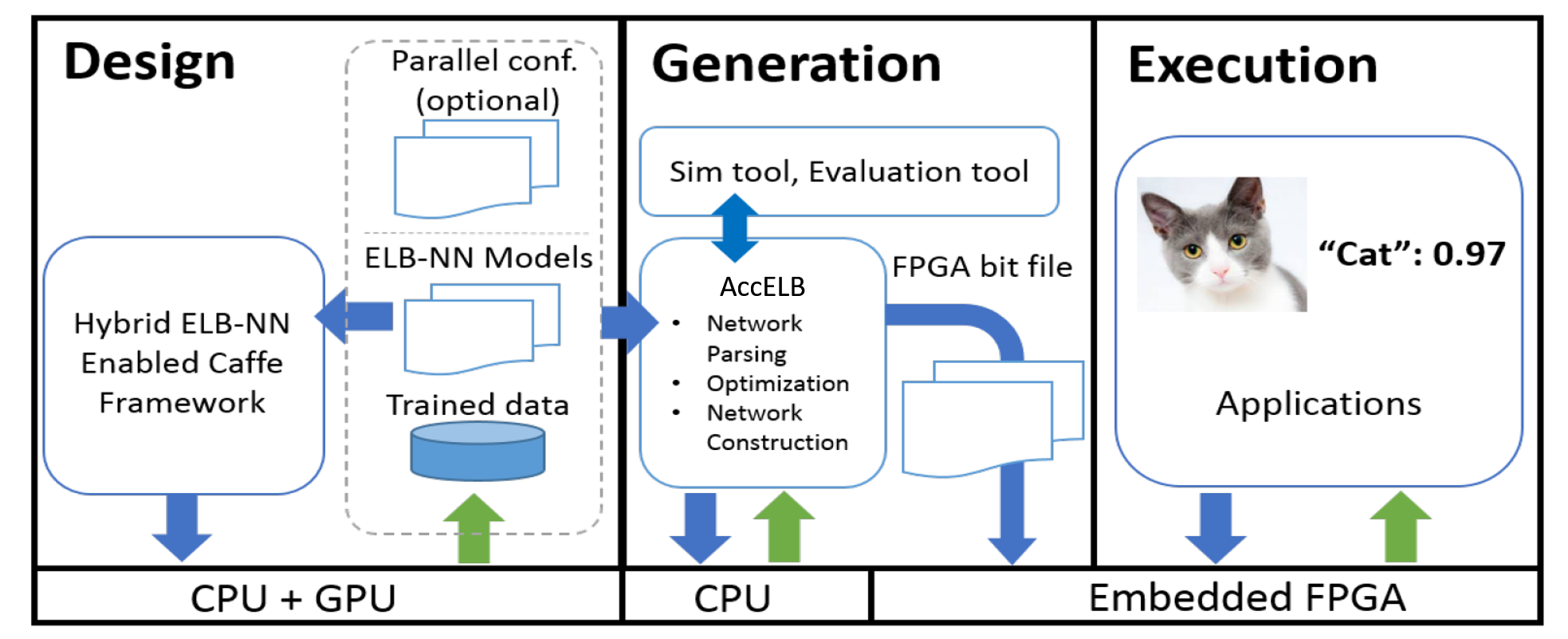}
  \caption{Hybrid ELB-NN design flow from network design, code generation to execution in FPGA.}
  \label{fig:overall_design_flow}
\end{figure}

To make hybrid ELB-NN design and performance evaluation in FPGA much easier, we build an end-to-end design flow from training to deployment in embedded FPGA. As shown in Fig. \ref{fig:overall_design_flow}, it integrates three steps as \textit{design}, \textit{generation}, and \textit{execution}.

\textbf{\textit{Design}}: We develop a Caffe based framework extended from Caffe-Ristretto \cite{gysel2016hardware} to support the training of hybrid ELB-NN, which allows different data bit-widths (including binary and ternary status) for inter- or intra- network layer optimization and delivers great possibilities to explore the tradeoffs among hardware performance and network accuracy. This framework also provides a hybrid quantization scheme for more stable training and faster convergence. The trained model (including the Caffe prototxt file and binary Caffe model file) could be directly used to generate Verilog code for FPGA deployment.

\textbf{\textit{Generation}}: After settling down with the network model, our tool AccELB then covers three critical procedures as 1) network parsing for building network structures on a layer by layer basis from abstract network definitions (e.g. Caffe prototxt files); 2) auto optimization for performing design space exploration under given FPGA devices; and 3) network construction for instantiating the pre-built optimized hardware libraries. In network parsing, network components such as CONV, pooling, and Batch Normalization (BN) are analyzed and corresponding parameters are extracted, such as kernel size, feature map size, type of layer, etc. We also concatenate the highly related network components, such as CONV, BN, and ReLU layer to flatten the targeted neural network and limit the number of network layer types which would be instantiated on FPGA. In terms of auto optimization, we try to balance each pipelined stage with considerations of network's computation complexity, available hardware logic, and memory resource constraints and eventually maximize the overall throughput performance. After that, optimized parallelism parameters are generated so that the pre-built hardware libraries are instantiated with the most appropriate configurations. 

\textbf{\textit{Execution}}: In the last execution step, the accelerator is generated by AccELB with unified FIFO-like I/O interfaces which can be easily instantiated in FPGAs for board level testing. 

ELN-NN designers are likely to try different configurations in the whole design space to find an optimal quantization that satisfies the accuracy and throughput requirements. It is time-consuming to generate the FPGA bit file and evaluate the performance in FPGA board for each try. Thus, we provide an evaluation tool that can profile the FPGA hardware utilization, such as LUTs, memory and bandwidth cost, etc, and provide the estimated throughput as well before touching the hardware, which could greatly shorten the exploration time.   

\section{Hybrid ELB-NN in Case Study}
In this section, we use a case study to explain the motivation of hybrid ELB-NN. Through this experiment, we also provide some insights and guidelines for the hybrid-ELB design, which also guides our hardware architecture design for AccELB described in Sec. \ref{AccELB}.

\subsection{Experimental Network Setup}
\vspace{-0pt}
As a starting point, we study the Convolutional Neural Network (CNN), and specifically focus on tuning the precisions of weights and activations, to investigate their impacts on the overall classification accuracy. Although our proposed hybrid ELB-NN and the corresponding automation tool AccELB support diverse precisions of the activations crossing layers, in this experiment, we still use unified precision for the intermediate activations to reduce the searching space. 

The binary weights can be calculated as Eq. \ref{eq:binary}. Here $\tilde{w}$ represents full precision weights after back propagation while {\bf $E(|\tilde{w}|)$} represents the mean of all the full-precision weights as a scaling factor. For the ternary training, the $w_t$ (representing ternary weight) can be calculated following Eq. \ref{eq:ternary}. Here we set the threshold $w_{thres}=0.7E(|\tilde{w}|)$ as suggested in \cite{Ternarynet}; we also follow \cite{Ternarynet} to calculate the scaling factor $E$. 
We use 8 bits fixed point in the first and the last layer's weights, where generally higher data precision is required than the intermediate layers. The image input (RGB data) and output (last FC's output) are also quantized to 8 bits and 16 bits respectively in our design.
\begin{equation}
w_b = sign(|\tilde{w}|)\times\textbf{E}(|\tilde{w}|)
\label{eq:binary}
\end{equation} 
\begin{equation}
w_t = 
\begin{cases}
sign(\tilde{w})\times $E$ & \mbox{$|w_t|>w_{thres}$}\\
0 & \mbox{$|w_t| \leq w_{thres}$}
\end{cases}
\label{eq:ternary}
\end{equation}



We modified Alexnet \cite{krizhevsky2012imagenet} with our hybrid ELB-NN method as the experimental benchmark. Alexnet has diverse CONV and FC layer structure and is very suitable for our investigation as a starting point. 
\subsection{Evaluation Results}
\label{training_accuracy}
\begin{figure}
    \begin{centering}
    \vspace{-0.1in} 
    \includegraphics[width=0.9\columnwidth]{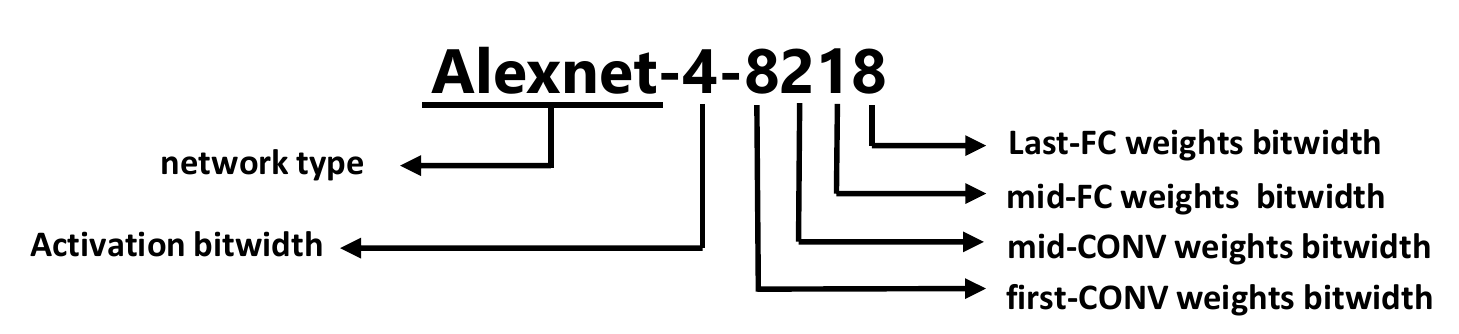}
    \caption{Network representation}    
    \vspace{+2pt}
    \label{fig:representation}
    \end{centering}
 \end{figure}

To sum up, we focus on studying the impact of {\bf (1)} the binary and ternary weights of the CONV layer and FC layer, and {\bf (2)} low bit-width activations. As we mentioned before, we apply relatively high precision (i.e., 8 bits) to the first CONV and last FC layer. We use mid-CONV to denote all the CONV layers except the first CONV layer, and use mid-FC to denote all the FC layers except the last FC layer. The naming rule of proposed hybrid precision network can be referred to Fig. \ref{fig:representation}. We first study a full 8 bits fix-point network (AlexNet-8-8888) with all weights and activations quantized to 8 bits. Then, we reduce the precision of the mid-CONV and mid-FC layers to binary or ternary, and study the accuracy characteristics. Finally, we reduce the bit-width of activations to investigate the impact. Our detailed experiments and the corresponding classification results are summarized in Table \ref{tbl:accuracy}.

\begin{table}
  \centering
  \caption{Classification accuracy for different weights and activation precision based ImageNet 2012 \cite{deng2009imagenet} dataset.}
  \label{tbl:accuracy}
  \begin{tabular}{|c|c|}
  \hline
    Network precision & Accuracy (Top-1)\\
    \hline
    Alexnet with float32 & 55.9\%\cite{dorefa} \\
    \hline
    Alexnet-8-8888 & 54.6\% \\
    \hline
    Alexnet-8-8228 & 53.3\% \\
     \hline
    Alexnet-8-8218 & 52.6\% \\
    \hline
    Alexnet-8-8118 & 51.1\% \\
    \hline
    Alexnet-4-8218 & 49.3\% \\
    \hline
    Alexnet-2-8218 & 46.1\% \\
    \hline
    Alexnet-4-8218 (w/o group) & 53.2\%\\
    \hline
    Alexnet-4-8218 (extended) & 54.5\%\\
    \hline
  \end{tabular}
  \vspace*{-8pt}
\end{table}

The preliminary results of the classification accuracy are encouraging. We find that the 8 bits fixed-point design (Alexnet-8-8888) only reduces the accuracy of the original network by 1.3\% after training. Further, with the precision of mid-CONV and mid-FC layer becoming ternary (Alexnet-8-8228), the classification accuracy reduces by another 1.3\% compared with the 8 bits version. Then, the accuracy reduces by an additional 0.7\% when the weights of the mid-FC layers are binarized (Alexnet-8-8218). That said, with the ternary and binary weights used in mid-CONV and mid-FC layer respectively, the accuracy is still promising. In other words, the network is relatively robust to the precision of weights. This insight is very helpful when mapping this quantized model to hardware, because the neural network acceleration usually suffers from the limited memory bandwidth, especially for the edge devices, and reducing the precision of weights to binary/ternary can reduce the required memory bandwidth dramatically. It is advisable to use binary status for the FC layer to seek even lower bandwidth consumption with very limited accuracy loss. More accuracy degradation of 1.5\% is observed when reducing the precision of CONV weights from ternary to binary status (AlexNet-8-8118), which indicates the weight precision of CONV layer has larger impact on accuracy than that of FC layer. 

The precision of activations also has more significant impact to the classification accuracy. Compared to the Alexnet-8-8218 in Table \ref{tbl:accuracy}, reductions of 3.3\% and 6.5\% accuracy are observed when the precision of activation moves from 8 bits to 4 bits (Alexnet-4-8218) and 2 bits (Alexnet-2-8218). It also gives us some ideas why fully binarized network suffers significant accuracy loss as stated in \cite{xnor-net}. In ELB-NN, each CONV or FC layer is followed by a Batch Normalization (BN) and a ReLU layer, which generates non-negative activations. It means the sign bit is useless and it could be used by the fractional part. For example, a 4 bits non-negative activation has the same representative capability as a 5 bits signed activation. Thus, it is a good choice to allocate all available bits to the value of activation instead of wasting one bit as a sign bit.

To further investigate the impact of model complexity in ELB-NN, we firstly disable the group function in CONV2, CONV4, CONV5 of the Alexnet, which is originally proposed to handle the GPU memory issue and distribute one CONV layer to two GPUs. As a result, the amount of computation is increased by 80\%, and a significant accuracy improvement of 3.9\% is observed. Secondly, starting from Alexnet-4-8218 (w/o group), we increase the kernel number for the five CONV layers from C96-C256-C384-C384-C256 to C128-C384-C512-C512-C384 to enlarge the model's representation capability, and 1.3\% accuracy gain is obtained by affording extra 61\% computation when comparing to Alexnet-4-8218 (w/o group). Our initial experimental results suggest that increasing the model complexity can definitely bring back the accuracy which is sacrificed by using ELB quantizations. Given the large design space, it is still challenging to find an optimal design in the entire design space in terms of various ELB settings, model scale and hardware efficiency. However, a heuristic method could be leveraged along with the design flow proposed in Sec. \ref{deign_overview} in an iterative way to find an optimized design. We may also leverage the machine learning based technology to perform fast design space exploration, which could be the future work.

\section{AccELB Design}
\label{AccELB}
The detailed procedure of RTL code generation using AccELB has been described in Sec. \ref{deign_overview}. In this section, we focus on its hardware architecture design.

\subsection{Data Access Pattern Investigation}
How to effectively utilize the bandwidth between the on-chip and off-chip memory is the key question when designing an accelerator. Most of the prior approaches perform the memory access in a recurrent manner, which means they reuse the same computation units and let all layers in DNNs share the same hardware accelerator with periodically data load/store operations. For each CONV layer, the computation engine needs to load the input feature map from the off-chip memory and store the output feature map back to the memory when the CONV computation completes. This transfer of feature map from and to external memory is costly in terms of latency and energy, which is much more serious for high resolution input. Take the classic Alexnet for the example, as mentioned in \cite{FuseCNN}, the feature map data occupies about 25\% of overall data in convolutional layers, and this number increases to over 50\% for more advanced models, such as VGG and GoogleNet. While in ELB-NN, more bits are required for feature maps comparing to weights as discussed in Sec. \ref{training_accuracy}, and the feature map movement becomes constrained by the bandwidth consumption. A typical configuration of 8 bits activations and ternary weights will make feature map data to be around 60\% and 80\% of the overall data for Alexnet and VGG, respectively. If considering high definition (HD, such as 720P/1080P) inputs, which are necessary for small object detection, nearly all the data movements relate to the intermediate feature maps. A layer fusion based CNN accelerator was proposed to significantly alleviate the burden of feature map data movement in \cite{FuseCNN} where output feature maps are not stored off-chip but kept inside the on-chip memory. In this paper, we use a full pipeline design and maintain the output feature maps in on-chip memory, which totally eliminates the tremendous intermediate feature map data moving cost. Meanwhile, the pipeline design also naturally supports the hybrid ELB-NN that requires diverse precision in layer granularity.

\subsection{Hardware Architecture}

\begin{figure}
    \begin{centering}
    \includegraphics[width=3.4in]{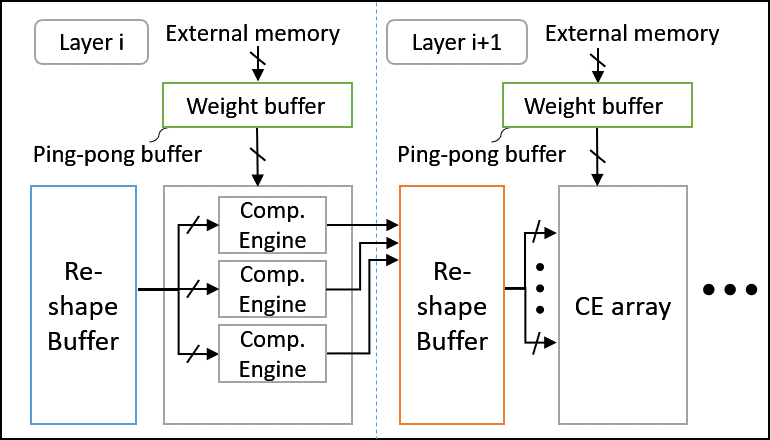}
    \caption{AccELB architecture}
    \label{fig:architecture}
    \vspace{-8pt} 
    \end{centering}
 \end{figure}
 
We apply a pipeline structure in AccELB to better accommodate high throughput and real-time performance. 
Each pipeline stage handles one concatenated layer (i.e., CONV + BN + ReLU) and video/image frames can be streamed in with very high throughput.  
%
Fig. \ref{fig:architecture} shows two of the proposed pipelined stages where data can be fetched from two sources as the FPGA on-chip memory (BRAM-implemented buffers) and the external memory (DRAM). The on-chip reshape buffers are built for storing input feature map, while the weight buffers between external memory and Computation Engine (CE) are set up as caches of the external DRAM for reducing data access latency. We design an array of CEs (CE array) for parallel computing. 
More details are described below.

\subsubsection{Computation Engine (CE)}
\begin{figure}[t]
    \begin{centering}
    \includegraphics[width=0.9\columnwidth]{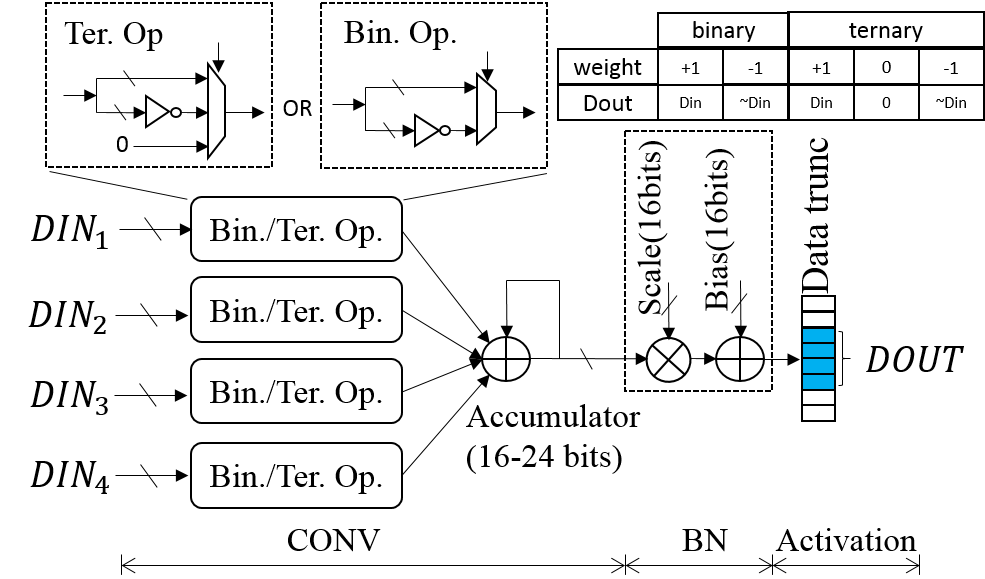}
    \caption{Computation engine (CE) with binary and ternary logic operations}
    \vspace{-10pt} 
    \label{fig:kernel}
    \end{centering}
 \end{figure}

Since the major part of convolutional computation is located in multiple nested loops, we propose a parameterized CE to introduce parallel instantiation by handling a certain number of CONV kernels and channels simultaneously. We also propose a flattened layer hierarchy operation by concatenating CONV, BN, and activation function and eventually improve the performance of the loops. Inside the CE, highly related network components, such as CONV, BN, and activation function, are concatenated to form an compact and efficient structure. As shown in Fig. \ref{fig:kernel}, we present a four-input-parallel CE as an example, where four inputs are first processed by ELB CONV (including binary/ternary operations and accumulation) and then followed by BN and activation function. 

In CONV, we follow the rules in the top-right of Fig. \ref{fig:kernel} and implement binary and ternary operations with \textit{combinatorial logic}. By using multiplexer, the output of Ternary or Binary operator can be selected among the input, the inverse of input, and zero accordingly. 
That said, no multiplications are required in CONV. The precision of the accumulator is adjustable to avoid overflow when large CONV kernel or higher precision of input data are applied. This adjustable precision is intended to allow more flexible quantization designs and maintain the output accuracy. For higher number of inputs, an adder tree will be used before the accumulator for easier timing enclosure. BN in the inference stage could be degenerated to a simple formula $\alpha x+\beta$ (according to the definition of BN), and the scaling factor $E$ for binary/ternary weights in Eq. \ref{eq:binary} and Eq. \ref{eq:ternary} can also be integrated to this formula, resulting a total scaling factor of $\alpha E$. The last component is the activation function, such as ReLU, in which the output is processed by a simple saturated truncation to a desired bit-width.



\subsubsection{Memory System}
In this section, the hierarchical memory system is introduced to tolerate the data access latency from external memory. Since there are few on-chip memory blocks available in embedded FPGAs, we need to limit the on-chip memory utilization by placing the trained weights off-chip especially for large DNNs. Although the weights and biases are quantized, the on-chip memory can be still insufficient when trying to deploy large-scale DNNs. To solve the data access latency, we insert on-chip ping-pong buffer between the CE and the external memory. This hierarchical memory system enables AccELB with the capability of deploying large scale models.


\section{Experimental results}

\begin{table*}[t]
\scriptsize
\vspace{-2pt}
\caption{AccELB performance evaluated on Xilinx ZC706 board}
\label{tab:performance}
\vspace{-6pt}
\begin{center}
\newcommand{\tabincell}[2]{\begin{tabular}{@{}#1@{}}#2\end{tabular}}
\begin{tabular}{|c|c|c|c|c|c|c|c|c|c|}
\hline
\multirow{2}{*}{Network} & \multicolumn{4}{c|} {Utilization (include SDK, around 11\% LUT and BRAM)}  & \multirow{2}{*}{\tabincell{c}{Batch\\size}} & \multirow{2}{*}{\tabincell{c}{Bandwidth\\(GBytes/s)}} & \multirow{2}{*}{\tabincell{c}{Complexity\\(GOP)}} & \multirow{2}{*}{\tabincell{c}{Speed\\(imges/s)}} & \multirow{2}{*}{\tabincell{c}{Performance\\(TOPS)}}\\ 
\cline{2-5} 
 & LUT  & FF  & BRAM & DSP  & \multicolumn{1}{c|}{} & \multicolumn{1}{c|}{} & \multicolumn{1}{c|}{} & \multicolumn{1}{c|}{} & \multicolumn{1}{c|}{} \\
\hline 
Alexnet-8-8888 (baseline) & \tabincell{c}{86262(39\%)} & \tabincell{c}{51387(12\%)} & \tabincell{c}{303(56\%)} & \tabincell{c}{808(90\%)} & 2 & 10.8 & 1.45 & 340 & 0.493 \\ 
\hline
Alexnet-8-8218 & \tabincell{c}{103505(47\%)}  & \tabincell{c}{90125(21\%)} & \tabincell{c}{498(91\%)} & \tabincell{c}{550(61\%)} & 5 & 3.35 & 1.45 & 856.1 & 1.24 \\
\hline
Alexnet-4-8218 & \tabincell{c}{105673(48\%)} & \tabincell{c}{94149(22\%)} & \tabincell{c}{463(85\%)} & \tabincell{c}{880(98\%)} & 8 & 3.35 & 1.45 & 1369.6 & 1.99 \\ 
\hline
Alexnet-4-8218 (w/o group) & \tabincell{c}{127393(58\%)}  & \tabincell{c}{105328(24\%)} & \tabincell{c}{435(80\%)} & \tabincell{c}{839(93\%)} & 7 & 4.30 & 2.61 & 1198.5 & 2.59 \\
\hline
Alexnet-4-8218 (extended) & \tabincell{c}{124317(57\%)}  & \tabincell{c}{101558(23\%)} & \tabincell{c}{481(88\%)} & \tabincell{c}{783(87\%)} & 7 & 3.4 & 4.22 & 599.2 & 2.53 \\
\hline
\hline
VGG16-4-8218 & \tabincell{c}{112992(52\%)}  & \tabincell{c}{99396(23\%)} & \tabincell{c}{509(93\%)} & \tabincell{c}{298(33\%)} & 2 & 5.85 & 31.0
 & 110.7 & 3.43 \\
\hline
VGG16-2-8118 & \tabincell{c}{137973(63\%)} & \tabincell{c}{113262(26\%)} & \tabincell{c}{499(92\%)} & \tabincell{c}{651(72\%)} & 3 & 6.67 & 31.0
 & 332.2 & 10.3 \\
\hline
\end{tabular}
\end{center}
\end{table*}

\begin{table*}[t]
\scriptsize
\vspace{-2pt}
\caption{Comparison with existing FPGA-based low precision DNN accelerators}
\vspace{-6pt}
\label{tab:comp}
\begin{center}
\newcommand{\tabincell}[2]{\begin{tabular}{@{}#1@{}}#2\end{tabular}}
\begin{tabular}{|c|c|c|c|c|c|c|}
\hline 
Reference & \cite{zhao2017accelerating} & \cite{umuroglu2017finn}  & \cite{Nicholas2017scale} & \cite{Nurvitadhi2016AcceleratingBN} & AccELB(1)-VGG16 & AccELB(2)-VGG16\\ 
\hline
FPGA chip & XC7Z020 & XC7Z045 & XCKU115 & Arria10 1150 & XC7Z045 & XC7Z045\\
\hline
Frequency & 143MHz & 200 MHz & 125 MHz & N/A & 200MHz & 200 MHz \\
\hline
Network Type & Binary & Binary & Binary & Binary & Hybrid(4-8218) & Hybrid(2-8118)\\
\hline
kLUTs (used/total) & 46.9/53.2 & 82.9/218.6 & 392.9/663 & N/A & 113.0/218.6 & 138.0/218.6\\
\hline
Performance (TOPS) & 0.21 & 9.1 & 14.8 & 9.8 & 3.43 & 10.3\\
\hline
efficiency (GOPS/kLUT) & 3.95 & 41.6 & 22.3 & N/A & 15.6 & 47.1\\
\hline
\end{tabular}
\vspace{-6pt}

\end{center}
\end{table*}

A Xilinx ZC706 evaluation platform with Zynq XC7Z045 FPGA (which consists of a Xilinx Kintex-7 FPGA and a dual ARM Cortex-A9 CPU) is used as the evaluation device. The hybrid ELB-NN with different precision configurations are automatically deployed and optimized by AccELB. After synthesis by Vivado 2016.4, placement and routing are completed subsequently showing resource consumptions and performance in Table \ref{tab:performance}. We achieved 200 MHz working frequency in this 28nm chip, which also shows the advantage of a full RTL level implementation. The SDK occupies around 11\% LUT and BRAM for generating the necessary infrastructure, such as the image/video data interface, the video/accelerator DMAs, the MIG, etc. 

\subsection{Alexnet Performance}
As shown in Table \ref{tab:performance}, Alexnet with 8 bits weight and activation implementation is firstly presented as a baseline. To balance the accuracy and off-chip memory access bandwidth, we use ternary weights for the mid-CONV layer and binary weights for mid-FC layer. In Alexnet, the computation of the first CONV layer occupies around 14.5\% of the total computation. 
We can achieve the throughput of 171.2 images/s without using batch. 

For Alexnet-8-8218, we can place at most 5 batches simultaneously and the BRAMs are first to be exhausted when placing more batches. While for Alexnet-4-8218, we can place 8 batches simultaneously and the DSP resource that is mostly consumed by the first layer becomes the bottleneck although we have explored to use one DSP block for two 8 bits multipliers \cite{XilinxInt8}. Therefore, we can not benefit more in terms of total throughput by further reducing the precision of activations to 2 bits. In our design, there are several multipliers for address computation in the pipeline control logic. In this case (Alexnet-4-8218), we use LUT resource to instantiate these address-computation multipliers instead of build-in DSP blocks since the performance is bounded by the DSP resource. Thus, the LUT usage in this case is a little higher than usual. The bandwidth requirements of both ELB settings are significantly reduced to 3.35 GBytes/s, which is acceptable for edge-devices with power constraints.

The Alexnet-4-8218 (w/o group) achieves the throughput of 1198.5 images/s, which is slightly slower than Alexnet-4-8218 (w/ group). However, the accuracy is significantly improved by 3.9\% as reported in Table \ref{tbl:accuracy}, which even outperforms Alexnet-8-8218 by 0.6\%. The speed of extended Alexnet-4-8218 can reach up to 599.2 images/s, which surpasses the baseline (Alexnet-8-8888) by 1.76x while still keeping the same accuracy as reported in Table \ref{tbl:accuracy}. As for the bandwidth cost, a significant reduction of 68\% is observed when comparing to the baseline resulting less power consumption. Meanwhile, bandwidth reduction can also reduce DDR hardware cost (e.g., using 32 bits instead of 64 bits DDR), since 3.4 GBytes/s bandwidth requirement can be well satisfied by using 32 bits DDR3 @1600MT/S instead of 64 bits.

\subsection{Scaling to large scale networks}
In order to demonstrate AccELB's capability of scaling to very large scale networks, we then present performance evaluation for VGG16 using ELB-NN, which is around 20x more complicated than standard Alexnet. 
We use VGG16-4-8218 as the case study. The top-1 accuracy is 55.8\%, which is trained using the same epochs and hyper parameters as those in Alexnet. We achieve the higher performance of 3.43 TOPS as shown in Table \ref{tab:performance}, which outperforms the best performance in Alexnet-4-8218 (w/o group) with the same ELB setting but using less LUT resources. The major reasons are {\bf (1)} The kernel number of CONV in VGG16 is much larger than that in Alexnet, and we can use higher parallelism in CE for VGG16. It will result in higher LUT efficiency, which is defined as GOPS$/$\# (of LUTs used). Based on our experiment on a typical CE with ternary weight and 4 bits data width, the LUT efficiency of 64 inputs outperforms that of 16 inputs by 1.4x. {\bf (2)} VGG16 has unified CONV kernel (3$\times$3 with stride 1) and pooling pattern (2$\times$2 with stride 2), which makes the whole pipeline perfectly balanced without any computation resource waste under the constraint that parallel factors in CEs should be power of 2. The more balanced for latency of each layer, the higher LUT efficiency. 

\subsection{Comparison to FPGA Accelerators}
We compare our design with four recent FPGA-based accelerators in Table \ref{tab:comp}. All of these four existing works focus on the BNNs, and only consider small models for their benchmark (even Alexnet was not considered). For their cases, all the binarized weights could be stored in the on-chip memory of FPGA. In our AccELB, we do not only consider the implementation of hybrid quantization, but also investigate the performance in very large scale models, such as VGG16. In this case, the weights can not be stored in on-chip memory any more even when all the weights are binarized, and a memory hierarchy mechanism should be involved. For fair comparison, we use the hybrid configuration of 2-8118, which is the closest precision setting to BNN. We achieve the peak performance of 10.3TOPS. Even we use one more bit for the activations, we still outperform the existing works except \cite{Nicholas2017scale}, in which they used a much more powerful FPGA with a greater amount of LUT resource (2x more than ours). However, for the LUT efficiency in terms of GOPS per kilo LUTs, our implementation still outperforms \cite{Nicholas2017scale} by 2x, and is better than the most efficient work in \cite{umuroglu2017finn} by 13\%.

\subsection{Comparison to GPU}

\begin{figure}[t]
  \centering
  \includegraphics[width=0.98\columnwidth]{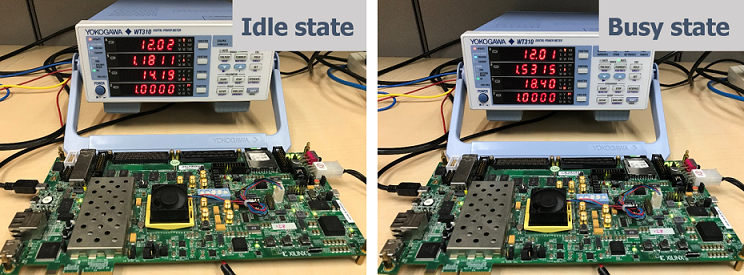}
  \caption{Board level implementation in Xilinx ZC706 and power analysis during (left) idle state and (right) busy state.}
  \vspace{+6pt}
  \label{fig:power}
\end{figure}

\begin{table}[t]
\scriptsize
\caption{Energy efficiency comparison (Alexnet)}
\vspace{-2pt}
\label{tb:comp_gpu_cpu}
\begin{center}
\newcommand{\tabincell}[2]{\begin{tabular}{@{}#1@{}}#2\end{tabular}}
\begin{tabular}{|c|c|c|c|c|c|}
\hline
Platform & Precision & \tabincell{c}{Batch\\size} & \tabincell{c}{Speed\\(img./s)} & \tabincell{c}{Power\\(w)} & \tabincell{c}{Efficiency\\(img./s/w)} \\ \hline
AccELB(1) & 4-8218 & 8 &  1369.6 & 4.2 & 325.3\\ \hline
AccELB(2) & 8-8218 & 5 & 856.1 & 4.1 & 208.8 \\ \hline
AccELB(3) & 4-8218 (extended) & 8 & 599.2 & 4.8 & 124.8 \\ \hline
GPU-TX2 & FP16 & 128 & 463 & 5.6 & 82.7\\ \hline
GPU-P4 & INT8 & 128 & 6084 & 56 & 109\\ \hline
GPU-P40 & INT8 & 128 & 11000 & 155 & 71\\ \hline
\end{tabular}
\end{center}
\vspace{-8pt}
\end{table}

We extend our comparison to latest GPUs both in data center \cite{Nvidia-P4-p40} and edge device\cite{NVidia_TX2} as shown in Table \ref{tb:comp_gpu_cpu}. 
All these hardware platforms are running inference for ImageNet-2012 on Alexnet. We use a Yokogawa WT310 digital power meter to measure the power consumption. As shown in Fig. \ref{fig:power}, the proposed design (Alexnet-4-8218) consumes 14.2W in idle state (when Linux OS is running at 800MHz) and 18.4W in busy state (when OS and the proposed design are running in ARM core and FPGA, respectively). To eliminate the affects of CPU (we did not use ARM core for DNN inference) and other unnecessary peripherals in this development board, we subtract the two power readings. Thus, the power consumed by target FPGA is 4.2W. 

As is clear from the comparison for the ELB-NN in Table \ref{tb:comp_gpu_cpu}, the energy efficiency in our proposed designs (Alexnet-4-8218 and Alexnet-8-8218) surpasses those of the latest Nvidia GPUs for data center (P4) and edge (TX2) inferences by up to 3.0x and 3.9x (the Alexnet-4-8218 case) even if we use much smaller batch size, but we also pay the price for the accuracy. Referring to Table \ref{tbl:accuracy}, 8-bit GPU's accuracy would be 54.6\%, where our implementations have accuracy of 52.6\% for AccELB(2) and 49.3\% for AccELB(1). A clear tradeoff between accuracy and efficiency can be observed. For the extended Alexnet-4-8218, which has the same accuracy as INT8 implementation, the energy efficiency still outperforms that of the most efficient GPU (P4) solution by 14\%.

GPU can also support ELB-NN by using different kernel implementations. However, existing literature \cite{Courbariaux-binary} suggests that BNN implementation (XNOR + popcount) is only 3.2x faster than FP32 cuBLAS implementation, and is very close to the INT8 performance \cite{INT8vsFP32}. Thus, no obvious gain is obtained in GPU when applying ELB-NN. While for FPGA, 4x throughput (Alexnet-4-8218) compared with the INT8 implementation (Alexnet-8-8888) is reported per Table \ref{tab:performance}.



\section{Conclusions}
In this paper, we proposed a design flow of accelerating large scale hybrid ELB-NN from training to deployment in FPGA to provide flexible accuracy and computation tradeoffs, which is critical for resource and power constrained edge devices. Especially, we presented AccELB, an end-to-end automation tool for deploying our proposed hybrid ELB-NN in embedded FPGAs to handle edge-computing applications with higher performance and energy efficiency. Provided with the trained ELB models, AccELB can generate a highly optimized FPGA implementation without involving any RTL programming. We tested AccELB on large scale ELB-NNs in a Xilinx ZC706 evaluation board and achieved peak performance up to 10.3 TOPS, as well as better energy efficiency (325.3 image/s/watt for running Alexnet) compared to existing approaches. With our proposed design flow from training to deployment, the user can easily bring the AI capability to edge-devices for fast prototyping or actual deployment with FPGAs.

\section*{Acknowledgment}
\vspace{-0pt}
This work was partly supported by the IBM-Illinois Center for Cognitive Computing System Research (C3SR) -- a research collaboration as part of IBM AI Horizons Network.

\bibliographystyle{unsrt}
\bibliography{main}

\end{document}